\documentclass[12pt]{iopart}

\usepackage{iopams} 
\expandafter\let\csname equation*\endcsname\relax
\expandafter\let\csname endequation*\endcsname\relax
\usepackage{amsmath}

\usepackage{hyperref}
\usepackage{caption}
\usepackage{todonotes} 
\usepackage{physics}

\begin{document}

\title[]{Computational Phase Transitions: Benchmarking Ising Machines and Quantum Optimisers}


\author{Hariphan Philathong} \ead{hariphan.philathong@skoltech.ru}
\author{Vishwa Akshay} \ead{akshay.vishwanathan@skoltech.ru} 
\author{Ksenia Samburskaya}\ead{k.samburskaya@skoltech.ru} 
\author{Jacob Biamonte} \ead{j.biamonte@skoltech.ru}

\address{Skolkovo Institute of Science and Technology\\ 30 Bolshoy Boulevard, Moscow 121205, Russian Federation}

\vspace{10pt}
\begin{indented}
\item[] \url{http://quantum.skoltech.ru} 
\item[] \today 
\end{indented}

\begin{abstract} While there are various approaches to benchmark physical processors, recent findings have focused on computational phase transitions. This is due to several factors.  Importantly, the hardest instances appear to be well-concentrated in a narrow region, with a control parameter allowing uniform random distributions of problem instances with similar computational challenge. It has been established that one could observe a computational phase transition in a distribution produced from coherent Ising machine(s). In terms of quantum approximate optimisation, the ability for the quantum algorithm to function depends critically on the ratio of a problems constraint to variable ratio (called {\it density}).  The critical density dependence on performance resulted in what was called, {\it reachability deficits}. In this perspective we recall the background needed to understand how to apply computational phase transitions in various bench-marking tasks and we survey several such contemporary findings. 
\end{abstract}

%
\noindent{\it Keywords}: QAOA, VQE, Ising Machines, Quantum Annealing, Graph Optimisers, Computational Phase Transitions.
%
\submitto{JPhys Complexity (invited perspective)}
%
%
%

\newpage

\section{Physics of computation and hardware benchmarking}

Physics has inspired many computational algorithms such as simulated annealing, Monte Carlo methods, random walks and more. These algorithms function by simulating a physical process---such as cooling---for the purpose of finding extrema (i.e.~minimisation, a.k.a.~optimisation). Recently such physics inspired algorithms have become increasingly replaced by actual physical devices. In other words, instead of simulating the physics inspiring a computer algorithm, one can build a physical computing machine that actually implements e.g.~the cooling process. Quantum effects are sought to accelerate the physical optimisation processes. 

Recent experimental progress has culminated to produce a range of physical optimisers. These optimisers essentially allow us to perform energy minimisation of what physicists call (tunable) Ising models. This is equivalent to what computer scientists call, quadratic binary optimisation. This approach is realized, for example, using {\it Ising machines} based on optical parametric oscillators \cite{inagaki2016coherent, mcmahon2016fully}, quantum annealers based on superconducting bistate systems \cite{venturelli2015quantum,weber2018hardware,chen2017progress,harris2018phase}, gain-dissipative platforms based on non-equilibrium condensates \cite{kalinin2018networks} and gate based superconducting electronics qubits executing quantum approximate optimisation (QAOA) \cite{willsch2020benchmarking,arute2020quantum}. One can also find quantum simulators and processors based on trapped ions \cite{islam2011onset} as well as optical approaches \cite{simon2011quantum, qiang2018large} and even few-monolayer spin ice films \cite{bovo2019phase}.  Approaches utilising silicon or diamond defects can also be found \cite{tosi2017silicon}. 

As each physical optimiser has been built and developed based on different technologies, these physical processors have their own advantages and disadvantages. D-Wave processors---based on superconducting circuits---utilise some quantum effects to find a minimum of a given objective function via a process of quantum annealing \cite{das2005quantum}. However, connectivity between binary units in this type of hardware is sparse and thus restricts the forms of objective functions to be implemented. On the other hand, coherent Ising machines \cite{inagaki2016coherent, mcmahon2016fully} provide all-to-all connectivity. Such approaches \cite{inagaki2016coherent, mcmahon2016fully} lack quantum effects. 

We do not know when these physical processors will outperform standard supercomputers. The average or general performance is not the only concern: indeed, what types of problem(s) might these physical processors perform well on? And will they perform well when standard numerical approaches fail? 

Although there are several avenues to benchmark physical processors, we have advocated heavily that one will consider computational phase transitions \cite{philathong2019computational,akshay2020reachability,akshay2020reachability-graph}. This is due to several factors.  Importantly the hardest instances appear to be well-concentrated in a narrow region, with a control parameter allowing uniform random distributions of problem instances which appear to pose similar computational challenge.  

In a recent theoretical proposal, several of the authors have sought to physically observe a computational phase transition in distributions produced from coherent Ising machine(s) \cite{philathong2019computational}. In addition, several of us have also applied these same tools to quantify the performance of QAOA \cite{akshay2020reachability, akshay2020reachability-graph}. Regarding QAOA, we found that the easy-to-hard transition plays an important role.  Namely, the ability for the quantum algorithm to function depends critically on the ratio of a problems constraint to variable ratio (called {\it density}).  The critical density dependence on performance resulted in what was called, {\it reachability deficits} \cite{akshay2020reachability, akshay2020reachability-graph}.

\section{Complexity, Physics and Phase Transitions}

In so-called complex systems, a system composed of interacting components, universally features phase transition(s). If suitable control parameters are varied, such systems undergo strong qualitative changes in its macroscopic properties. The characteristic descriptors (e.g.~the so-called critical exponents) are  shared by an abundance of physical systems \cite{binney1992theory}. Such descriptors exhibit discontinuities (at the transition point) in the large system size limit.

One of the best known model which exhibits phase transitions is Ising's model of magnetism. A sudden change in the behaviour of 2D the Ising model takes place at a critical temperature $T_{c}$ (the so-called Curie temperature). For $T \ll T_{c}$, the spins are aligned. For $T>T_{c}$, the magnetization is zero as a consequence of a random spin distribution. For $T=T_{c}$, the spin distribution appears to be clustered in a non-trivial way \cite{sole1996phase}. More details and a list of very different complex systems undergoing phase transitions have been investigated in \cite{sole1996phase}.

In quantum systems, the phase transition from one state to another is provided by adjusting a tuning parameter other than temperature \cite{sachdev2007quantum}. In the quantum Ising model, the phase transition takes place when the strength of transverse field is approximately the same as the magnitude of coupling interactions. In this regime, it is known that the model exhibits a complicated structure \cite{suzuki2012quantum}. More detailed discussions on phase transitions of transverse Ising models and the techniques developed for the study of critical behaviour are available in \cite{suzuki2012quantum,sachdev2007quantum}.

In computer science, so-called computational (or algorithmic) phase transitions have been studied extensively. The phenomena was first reported in the satisfiability ({\sf SAT}) problem \cite{crawford1996experimental,friedgut1999sharp,selman1996critical,selman1996generating,chvatal1992mick,goerdt1996threshold}, which is an {\sf NP}-complete decision problem \cite{cook1971complexity}. {\sf SAT} consists of determining whether a Boolean formula can evaluate to {\sf TRUE}, and hence be {\it satisfied}. A SAT formula in  conjunctive normal form (CNF) is a conjunction (Logical OR) of clauses where each clause is a disjunction (logical AND) of Boolean variables (or their negation). If $k$ is the number of variables in each clause, the problem is called $k$-{\sf SAT}. For example, consider a $3$-SAT formula of $5$ variables and $3$ clauses 
\begin{equation}
(x_{1} \lor \neg x_{2} \lor x_{3}) \land  ( \neg x_{1} \lor x_{4} \lor \neg x_{5})  \land  ( x_{2} \lor x_{3} \lor \neg x_{4}).
\end{equation}
The formula is satisfiable, as $x_1=1,\,x_2=1,\,x_3=0,\,x_4=1,\,x_5=0$. The density of variables to clauses is $\tfrac{3}{5}$. 

\begin{figure}[htb]
    \centering
    \includegraphics[width=\textwidth]{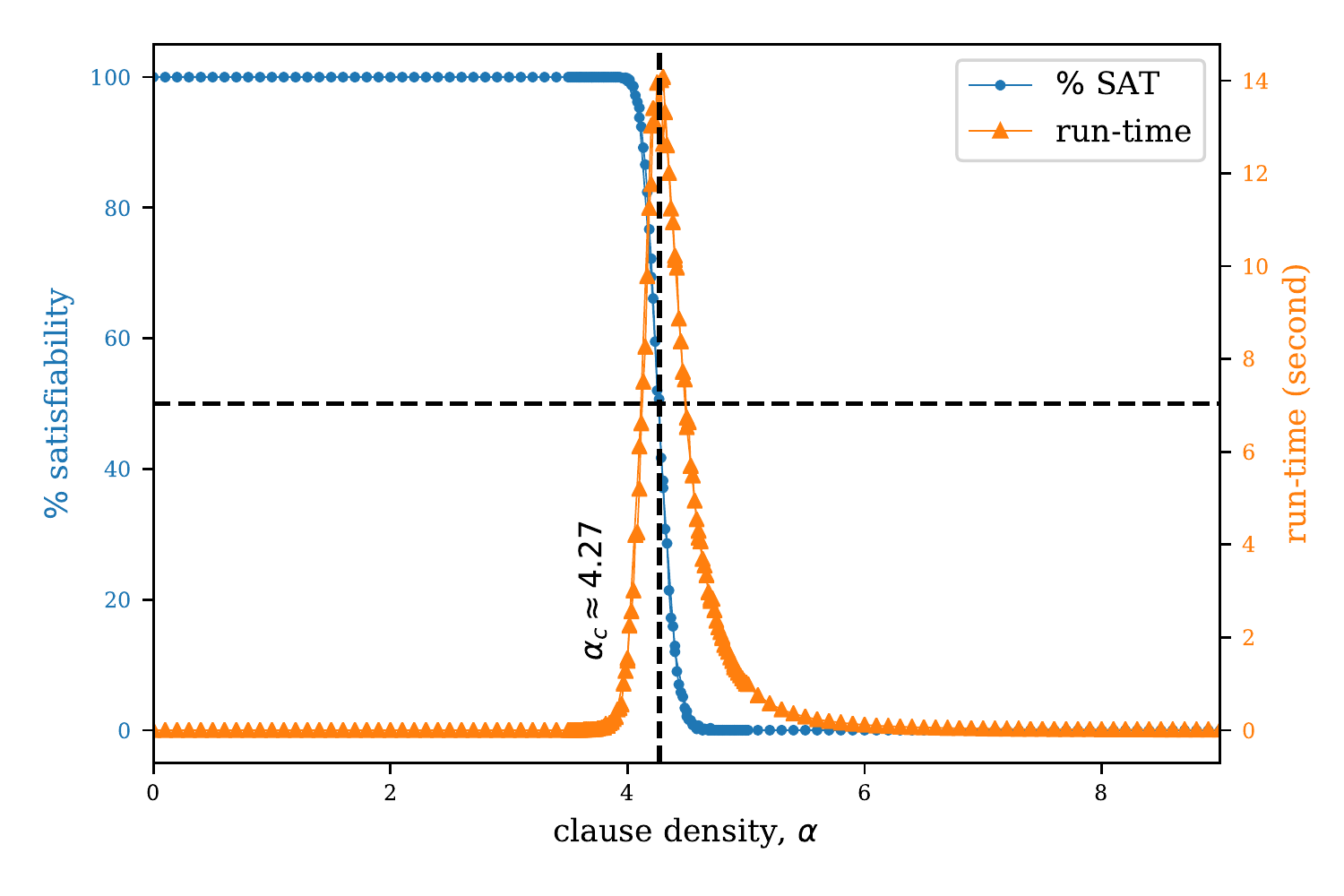}
    \caption{Percent of satisfiable instances (left axis) and average run-time (right axis) versus clause density, $\alpha$. For each data point, we randomly generated $1,000$ $3$-SAT instances with $300$ variables and made use of a so-called PycoSat in Python 3 which is a SAT solver based on a backtracking search algorithm \cite{biere2008picosat} to solve the problem instances with observed $\alpha_{c} \approx 4.27$. Figure reproduced from \cite{philathong2019computational} and based on works appearing in \cite{crawford1996experimental,friedgut1999sharp,selman1996critical,selman1996generating,chvatal1992mick,goerdt1996threshold}.}
\label{fig:PhaseTransitionSAT}
\end{figure}

It has been observed that the rapid change from satisfiable instances to unsatisfiable instances takes place at a certain value of an order parameter: clause to variable ratio (or clause density). Also it has been empirically observed that computational algorithms seem to slow down at this critical point (see Fig.~\ref{fig:PhaseTransitionSAT}). The requirement of increasing computational resources to solve instances suggests that difficult instances for the $k$-{\sf SAT} problem concentrate around this critical density (an easy-hard-easy transition) \cite{crawford1996experimental,friedgut1999sharp,selman1996critical,selman1996generating,chvatal1992mick,goerdt1996threshold}. These two features indeed are the signature of the computational (or algorithmic) phase transition. 

In addition to the {\sf SAT} problem, there are many more examples which admit computational phase transitions, such as scheduling \cite{bauke2003phase,beck1997constrainedness,herroelen1999phase,rieffel2014parametrized}, graph colouring \cite{boettcher2002optimization,boettcher2003extremal,boettcher2004extremal,cheeseman1991really,culberson2001frozen}, partitioning \cite{boettcher1999extremal,boettcher1999extremalgraphpartitioning,boettcher2000nature,boettcher2001extremal}, and the travelling salesman problem \cite{gent1996tsp, macready1996criticality,percus1999stochastic,zhang1996study,zhang2004phase}.

These computational phase transitions seem to be a priori different; in physical sciences, one is the onset of non-trivial macroscopic collective behavior in a system composed of a large number of elements that follow simple microscopic laws \cite{martin2001statistical}, and in computer science, one is computational difficulty existing in a solution search process. However computational difficulty is reflected in both solving problem instances and also simulating physical processes across such critical points \cite{de1999simulation, gomez2019review}.

The connection between computational complexity and phase transitions have been studied \cite{kirkpatrick1994critical, achlioptas2005rigorous, anderson1999solving, gomes2002satisfied,gomes2005can,monasson1999determining, mezard2002analytic, mezard2003passing,selman2008hard}. By this connection, randomly generated $k$-{\sf SAT} instances can be interpreted as random realizations of a spin glass written in the form of the generalized Ising Hamiltonian with at most $k$-body interactions. This correspondence has allowed physicists and computer scientists to analyze the critical behavior of $k$-{\sf SAT} in the language of statistical mechanics. It also has provided some logical basis for considering $k$-{\sf SAT} as a physical spin system capable of exhibiting phase transitions.

\section{Benchmarking physical Ising machines via Gibbs sampling}

So far, the algorithmic phase transition (the easy-hard-easy transition) has not been directly observed in contemporary physical computing devices which solve problem instances via physical means, such as annealers \cite{kirkpatrick1983optimization}, Ising machines \cite{inagaki2016coherent, mcmahon2016fully, abrams2019implementation}, and quantum enhanced annealers \cite{venturelli2015quantum, harris2018phase, johnson2011quantum, barends2016digitized, harris2010experimental,king2018observation}.

However, as reported in \cite{philathong2019computational}, numerical prediction shows that the signature of algorithmic phase transition is possible to be observed in such contemporary physical computing devices via Gibbs sampling. Considering the thermal Gibbs states of physical systems, the probability of sampling ground states decreases around the phase transition point. The authors \cite{philathong2019computational} report the difficulties in Gibbs sampling solutions of the {\sf SAT} problems are around the critical point.

This prediction connects the computational phase transition with physical processes. Since recent Ising devices \cite{inagaki2016coherent, mcmahon2016fully} and quantum annealers \cite{venturelli2015quantum,weber2018hardware,chen2017progress} have been constructed with increasing programability, the computational phase transition signature only now has the potential to be physically observed. As difficult problem instances are thought to concentrate at the transition point, computational phase transitions then provide a tool to benchmark such computing devices.

\section{Benchmarking NISQ processors via QAOA}

Variational hybrid quantum/classical algorithms have been developed to utilize noisy intermediate-scale quantum (NISQ) devices to find approximate solutions to combinatorial optimization problems. These variational algorithms---such as the variational quantum eigensolver (VQE)---and the quantum approximate optimization algorithm (QAOA) train parameterized quantum circuits by using measurement-feedback loops to optimize a given objective function. This approach has gained wide interest.  QAOA is now the most studied gate based model for optimization on NISQ architectures, due to its relative ease in realization on existing processors \cite{willsch2020benchmarking,arute2020quantum,qiang2018large,abrams2019implementation,pagano2019quantum,otterbach2017unsupervised} and also due to recent universality results \cite{morales2020universality,lloyd2018quantum}.

{\sf MAX-SAT} or maximum satisfiability, which is the optimization version (NP-hard) of decision {\sf SAT}, also features the computational phase transition and therefore fits our purpose of benchmarking quantum processors via QAOA. Unlike decision {\sf SAT}, {\sf MAX-SAT} focuses on finding variable assignments that maximize the number of satisfied clauses in a given instance. This problem exhibits an easy-hard transition at a critical clause density, instead of an easy-hard-easy transition like in decision SAT.

In \cite{zhang2001phase}, the well-known Davis-Putman-Loveland (DPL) algorithm, a backtracking method with unit resolution \cite{davis1962machine}, was used to study the phase transition of {\sf MAX $3$-SAT}. The fraction of literals with fixed values in all optimal solutions was considered as a so-called backbone, which corresponds to critically constrained variables. At the critical point, {\sf MAX-SAT} undergoes a sharp transition from nonexistence of backbones when under-constrained to large backbones when over-constrained. Moreover, in random graph theory description, the transition is generally related to the birth of a giant component \cite{coppersmith2004random}.

\begin{figure}[htb]
  \centering
  \includegraphics[width=0.75\textwidth]{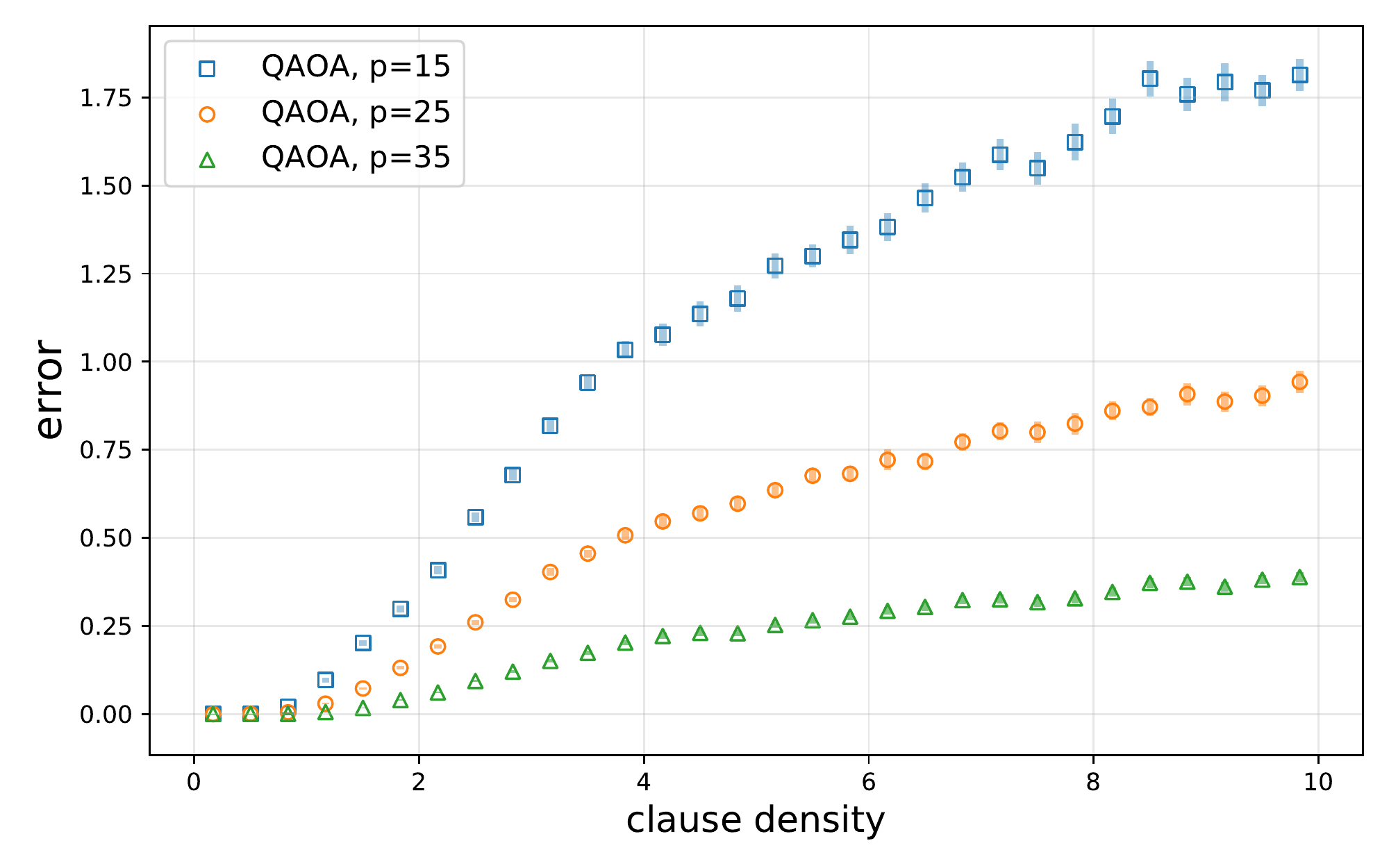}
  \includegraphics[width=0.75\textwidth]{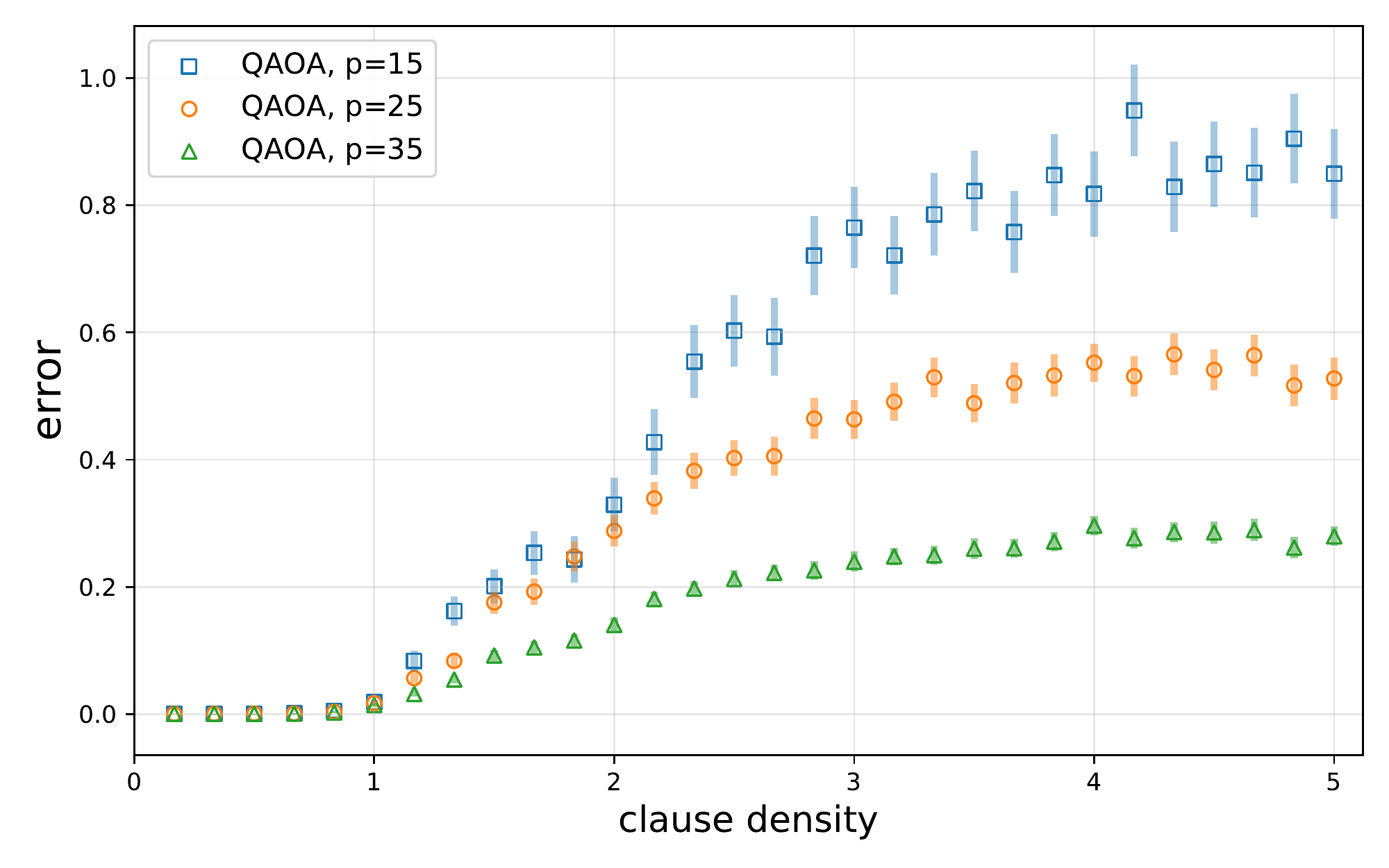}
\caption{Average error in QAOA approximation (error) vs clause density for MAX $3$-SAT (top) and MAX $2$-SAT (bottom) for differing QAOA depths. The $error = \min_{\psi \in \Omega_p} \bra{\psi}\mathcal{V}\ket{\psi} - \min_{\phi \in \mathcal{H}} \bra{\phi}\mathcal{V}\ket{\phi}$, where the first term indicates minimization of the objective function over the reachable state space $\Omega_{p} = \underset{\boldsymbol{\gamma},\boldsymbol{\beta}}{\bigcup} \{ \ket{\psi(\boldsymbol{\gamma},\boldsymbol{\beta})} = \prod_{k=1}^{p} \mathcal{U}(\gamma_{k},\beta_{k})\ket{+}^{\otimes{n}}\} \subseteq \mathcal{H}$ generated by the $p$-depth QAOA ansatzs whereas the second term represents the exact optimum computed by minimization over the entire Hilbert space $\mathcal{H}$. A Hamiltonian $\mathcal{V}^{\dagger}=\mathcal{V} \geq 0$ encodes solutions of a MAX-SAT instance into its ground state space, and the ground state energy is equal to minimal number of unsatisfiable clauses of the instance. Note that the error transitions occur at ${{\alpha }_{c}} \approx 1$. Figure reproduced from \cite{akshay2020reachability}.}
\label{fig:QAOAperformance}
\end{figure}

Reported in \cite{akshay2020reachability}, the authors numerically studied QAOA performances on the {\sf MAX-SAT} problems and calculated the average error in QAOA approximation (error) which is the difference between the optimum recovered from QAOA and the exact optimum computed via brute-force. The authors found that the error starts to increase beyond the critical clause density. This suggests that the QAOA is sensitive to this essential problem instance feature (see Fig.~\ref{fig:QAOAperformance}), brought about by {\it reachability deficits}.

Since the noisy intermediate-scale quantum (NISQ) devices have been recently proposed as a potentially viable application of quantum computers, the benchmarking problem of these devices is still open. Here we propose to use computational phase transitions as a tool to benchmark these computing devices via QAQA. We provide some numerical QAOA simulations on the MAX-SAT \cite{akshay2020reachability} and graph minimisation problems \cite{akshay2020reachability-graph} that have the potential to be physically realised in actual NISQ devices in the near future. Based on the numerical results, QAOA's performance is sensitive to these easy-hard transitions. As the difficulty of solving problem instances appears to relate to computational phase transitions and NISQ devices show increasing potential on implementation of QAOA \cite{willsch2020benchmarking,arute2020quantum,pagano2019quantum}, we can make use of these facts to benchmark optimisers.

\section{Conclusions}

Computational problems such as Boolean satisfiability feature abrupt changes similar to the phase transitions that are exhibited in condensed matter systems. From a computational standpoint, it is at these critical regions that a worst-case scaling in computational resources is observed, making them more difficult to solve. 

Recently---with the advancements made in physical computing devices---where the aim is to allow physical devices to naturally solve computational problem instances, If instances at critical regions are difficult to solve computationally, we expect the hardness be reflected in physical computing devices. It is under this notion we propose to use computational phase transitions as a tool to benchmark different physical computing devices \cite{philathong2019computational,akshay2020reachability,akshay2020reachability-graph}.

\section*{Acknowledgements}  
The authors acknowledge support from the research project, {\it Leading Research Center on Quantum Computing} (agreement No.~014/20).\\~\\
{\bf Competing interests:} The authors declare no competing interests.\\~\\
{\bf Data availability:} Any data that support the findings of this study are included within the article.\\~\\
{\bf Code availability:} The code for generating the data will be made available on GitHub after this paper is published.\\~\\

\bibliographystyle{unsrt.bst}
\bibliography{ref}

\end{document}